\documentclass[prl,aps,twocolumn,showpacs,superscriptaddress,floatfix,amsmath,amssymb]{revtex4}

\usepackage{graphicx}
\usepackage{dcolumn}
\usepackage{bm}
\usepackage[dvipdfm,bookmarks=true,colorlinks,%
            citecolor=blue,linkcolor=blue, hypertex, %
            breaklinks=true]{hyperref}

\begin{document}

\title{Partial order and finite-temperature phase transitions in Potts
models on irregular lattices}

\author{Q. N. Chen}
\affiliation{Institute of Theoretical Physics, Chinese Academy of Sciences,
P.O. Box 2735, Beijing 100190, China}

\author{M. P. Qin}
\affiliation{Institute of Physics, Chinese Academy of Sciences, P.O. Box
603, Beijing 100190, China}

\author{J. Chen}
\affiliation{Institute of Physics, Chinese Academy of Sciences, P.O. Box
603, Beijing 100190, China}

\author{Z. C. Wei}
\affiliation{Institute of Physics, Chinese Academy of Sciences, P.O. Box
603, Beijing 100190, China}

\author{H. H. Zhao}
\affiliation{Institute of Physics, Chinese Academy of Sciences, P.O. Box
603, Beijing 100190, China}

\author{B. Normand}
\affiliation{Department of Physics, Renmin University of China, Beijing
100872, China}

\author{T. Xiang}
\affiliation{Institute of Theoretical Physics, Chinese Academy of Sciences,
P.O. Box 2735, Beijing 100190, China}
\affiliation{Institute of Physics, Chinese Academy of Sciences, P.O. Box
603, Beijing 100190, China}

\date{\today}

\begin{abstract}
We evaluate the thermodynamic properties of the 4-state antiferromagnetic
Potts model on the Union-Jack lattice using tensor-based numerical methods.
We present strong evidence for a previously unknown, ``entropy-driven,''
finite-temperature phase transition to a partially ordered state. From the
thermodynamics of Potts models on the diced and centered diced lattices,
we propose that finite-temperature transitions and partially ordered
states are ubiquitous on irregular lattices.
\end{abstract}


\maketitle

The $q$-state Potts model has an essential role in the theory of
classical critical phenomena and phase transitions \cite{F.Y.Wu}.
Antiferromagnetic Potts models are rich and complex, displaying many
different types of behavior as a function of $q$ and of lattice
geometry, and the search for guiding principles continues. Key
questions include whether the model has a phase transition, if
this occurs at finite temperature, the nature of the low-temperature
phase, and the universality class of the transition.

In the Landau theory of phase transitions, these are described by an order
parameter. Minimizing the energy at low temperatures requires the order
parameter to be finite everywhere. However, in systems with extensive
ground-state degeneracy \cite{non0entropy}, of which the best-known example
is ice \cite{Pauling}, the nonzero entropy at zero temperature may cause a
different type of ordered state. A partial order, involving only some of the
lattice sites, is stabilized by minimization of energy in combination with
maximization of entropy. This ``entropy-driven'' transition can occur at a
finite temperature. Such partial order is known in both frustrated
\cite{spin-S,diamond} and unfrustrated \cite{x} systems. In the
latter case, partial order arises purely due to configurational entropy
effects, and the $q = 3$ antiferromagnetic Potts model on the diced lattice
provides an excellent example of the associated phase transition \cite{Diced}.

In this Letter we pursue the physical origin of the finite-temperature
phase transition in two-dimensional (2D) $q$-state Potts models. Although
these systems have no exact solution for $q > 2$, we employ tensor-based
numerical methods to obtain hitherto unavailable thermodynamic quantities.
We show that the $q = 4$ Potts model on the Union-Jack lattice exhibits a
finite-temperature transition to a state of partial order. We characterize
this transition by computing the entropy, specific heat, and magnetization,
quantities we also use to provide a complete discussion of the $q = 3$ model
on the diced lattice. We propose that finite-temperature transitions and
partially ordered states are a general property of Potts models on irregular
lattices.

\begin{figure}[t]
\includegraphics[width=8.5cm]{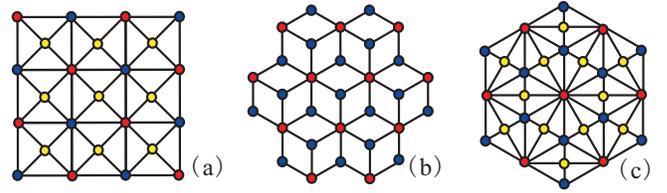}
\caption{(Color online) (a) Union-Jack lattice. Sites in sublattices A (red
circles) and B (blue) have coordination numbers $z_A = z_B = 8$, while those
in sublattice C (yellow) have $z_C = 4$. (b) Diced lattice. A sites (red)
have $z_A = 6$, while B sites (blue) have $z_B = 3$. (c) Centered diced
lattice: $z_A = 12$ (red), $z_B = 6$ (blue), and $z_C = 4$ (yellow).}
\label{ulpmf1}
\end{figure}

Let the $q$ states of the Potts model for lattice site $i$ be labeled
$\sigma_i = 0, 1, ..., q-1$. In the Hamiltonian,
\begin{equation}
{\cal H} = J \sum_{\langle i,j \rangle} \delta_{\sigma _{i} \sigma _{j}} -
H \sum_{i} \delta_{\sigma _{i},0}, \label{ehpm}
\end{equation}
$J > 0$ corresponds to the antiferromagnetic case and a field $H$ is coupled
to one of the $q$ states. We consider only a single coupling $J$ on every
bond, and begin with $H = 0$. We employ tensor-based numerical techniques
developed recently by a number of authors \cite{TRG,Jiang,Xie,Zhao,Vidal}
to compute the partition function $Z$ to high accuracy, and hence to obtain
the required thermodynamic quantities.

We illustrate the methods by focusing on the Union-Jack lattice [Fig.~1(a)].
The partition function
\begin{equation}
Z = {\rm Tr} e^{-\beta {\cal H}} = \sum_{\left\{ \sigma_{i} \right\}} \prod_{\square}
e^{-\beta {\cal H}_{\square }},
\end{equation}
$\square$ represents the structural unit (Fig.~2(a)) and
\begin{eqnarray}
{\cal H}_{\square }^{\rm UJ} = J \left( \delta_{\sigma_{1} \sigma_{2}} +
\delta_{\sigma_{2} \sigma_{3}} + \delta_{\sigma_{3} \sigma_{4}} +
\delta_{\sigma_{4}\sigma_{1}} \right) /2 \nonumber \\
 + J \left( \delta_{\sigma_{1} \sigma_{5}} + \delta_{\sigma_{2} \sigma_{5}}
 + \delta_{\sigma_{3} \sigma_{5}} + \delta_{\sigma_{4} \sigma_{5}} \right).
\label{ehsv}
\end{eqnarray}
The tensor $T_{\square} = e^{-\beta {\cal H}_{\square}}$ is defined \cite{TRG,Zhao}
by summing over the C-sublattice sites $\sigma_5$ and introducing the bond
(or dual-lattice) variables $\alpha = {\rm mod} (\sigma_1 - \sigma_2,q)$,
$\beta = {\rm mod} (\sigma_2 - \sigma_3, q)$, $\gamma = {\rm mod} (\sigma_3
 - \sigma_4, q)$, and $\delta = {\rm mod} (\sigma_4 - \sigma_1,q)$
[Fig.~2(a)]. The partition function
\begin{equation}
Z = \sum_{\alpha \beta \gamma \delta \dots} T_{\alpha\beta\gamma\delta}
T_{\alpha\epsilon\mu\nu} \dots
\label{epft}
\end{equation}
becomes a product of tensors defined on the square lattice [Fig.~2(b)].
The tensor renormalization group (TRG) \cite{TRG,Xie} is a real-space
coarse-graining method, in which the precision of the tensor contraction
is greatly enhanced by simultaneous renormalization of an ``environment''
block \cite{Xie,Zhao}. Infinite time-evolving block-decimation (iTEBD)
\cite{Vidal} is a projection method, in which projecting the transfer
matrix sufficiently many times on a random vector gives very accurately its
largest eigenvalue. Both methods simulate the properties of an infinite
system, the truncation being performed in the size $D$ of the rank-4 tensor
$T_{\alpha\beta\gamma\delta}$, which thus determines the precision. With an
accurate partition function, we then obtain all other thermodynamic
information.

\begin{figure}[t]
\includegraphics[width=7cm]{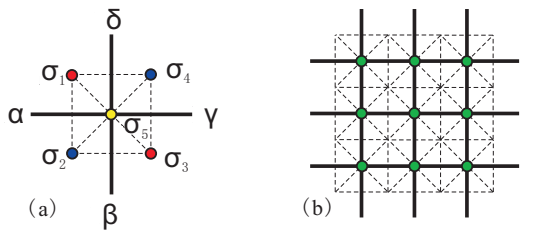}
\caption{(a) Schematic representation of the tensor $T$, obtained by
integrating over the C-sublattice sites and performing the dual
transformation. (b) The Union-Jack lattice (dashed lines) is transformed
to the dual square lattice (solid lines) on which the tensor $T$ is defined.}
\label{ulpmf2}
\end{figure}

One guiding principle for the antiferromagnetic Potts model in 2D concerns
the existence of a critical $q$, $q_c ({\cal L})$, for each lattice geometry
${\cal L}$. At zero temperature, neighboring sites may not have the same
``color'' $\sigma_i$ (Fig.~1), making the model equivalent to a vertex
coloring problem. By exploiting the Dobrushin Uniqueness Theorem
\cite{Dobrushin}, Salas and Sokal proved \cite{SS1} that the correlation
function decays exponentially at all temperatures (including zero)
for sufficiently large $q$, meaning a disordered ground state and no
phase transition. For sufficiently small $q$, ordered ground states
usually exist \cite{Diced,SS2}. For $q = q_c ({\cal L})$, $T = 0$ is a
critical point; this situation is common for many integer $q$ values,
including on the square \cite{SS2}, kagome \cite{kagome} (both $q_c = 3$),
 and triangular ($q_c = 4$) lattices \cite{triangular}.

In terms of $q_c$, the diced lattice [Fig.~1(b)] is anomalous. Despite
an average coordination number of 4, the model with $q = 3$ has a
finite-temperature phase transition, hence $q_c > 3$. This transition
is driven \cite{Diced} by the entropy available from the irregular nature
of the lattice, meaning that there are more sites of one type than of
others. The low-temperature phase can be ordered on the A sublattice (for
example, $\sigma_{i} = 0$) but not on B ($\sigma_{i} = 1$ or 2), creating
a partially ordered state. We investigate the hypothesis that such behavior
is generic for irregular lattices. We do this both by seeking an explicit
new example of a finite-temperature transition, which leads us to consider
the Union-Jack lattice, and by analyzing in detail the thermodynamic
properties of irregular lattices.

\begin{figure}[t]
\includegraphics[width=8.1cm]{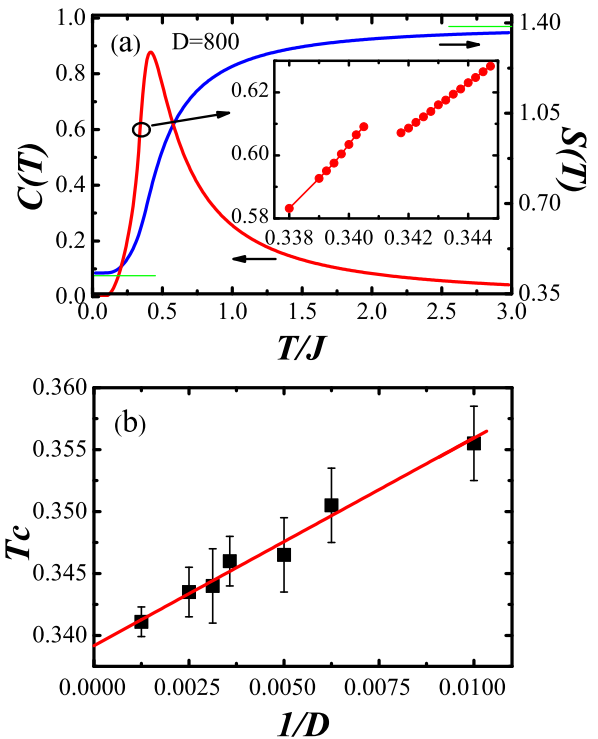}
\caption{(color online) (a) Entropy and specific heat for the $q = 4$ Potts
model on the Union-Jack lattice, computed with $D = 800$. Thin, green lines
denote the zero- and infinite-temperature limits of the entropy. Inset:
specific-heat discontinuity near $T_c$. (b) Linear fit of $T_c$ with $1/D$,
which extrapolates to $T_c(\infty) = 0.339(1)$. The error bar denotes
the upper and lower temperatures at the discontinuity.}
\label{ulpmf3}
\end{figure}

We begin by considering the $q = 4$ Potts model on the Union-Jack lattice.
This centered square lattice is tripartite [Fig.~1(a)]. While the $q = 2$
(Ising) model shows partial order and the $q = 3$ model is perfectly ordered,
the $q = 4$ model may be expected from its average coordination ($z = 6$)
to have zero-temperature order. To address the question of a phase transition
occurring instead at finite temperature due the irregular nature of the
lattice, we focus directly on the specific heat, shown in Fig.~3(a). The
small but clear gap in the curve [inset, Fig.~3(a)] indicates a discontinuous
second derivative of the free energy, and hence a second-order phase
transition. We have performed detailed calculations to delineate the nature
of the discontinuity over a range of $D$ values, and exploit the linear
scaling behavior of the matrix-product state with $1/D$ \cite{Scaling}, shown
in Fig.~3(b), to obtain the precise result $T_c = 0.339(1)$. Thus the $q = 4$
model on the Union-Jack lattice does indeed possess a finite-temperature
phase transition, becoming only the second known Potts model in this category.

This transition was to our knowledge neither known beforehand nor even
predicted. It underlines directly the importance of the irregular nature
of a 2D lattice in promoting a finite zero-temperature entropy and partial
order in the ground state. Our result is connected to several other models
in statistical mechanics. First, some Potts models may be mapped to a height
model, and when this mapping exists the system is critical or ordered at
zero temperature \cite{Diced,SS2}. As expected from our result, such a height
map does exist for the $q = 4$ Union-Jack lattice \cite{triangular}.
Second, at zero temperature, the $q = 4$ Potts model on the Union-Jack
lattice can be mapped directly to the 3-bond coloring problem on its dual,
the 4-8 lattice \cite{triangular}. The total number of states on the 4-8
lattice is known \cite{W4-8}, and hence the zero-temperature entropy of
the Union-Jack lattice should be $S^{\rm UJ} (0) = 2 \ln W_{4-8} = 0.430997$,
with $W_{4-8} = 1.24048$. Our numerical result is $0.430999$ [Fig.~3(a)].

Third, the bond-coloring problem on the 4-8 lattice is further equivalent
to a fully-packed loop (FPL) model (on the same lattice), obtained by
considering all configurations of noncrossing closed loops. The FPL partition
function is $Z = \sum_{G} n^{N_{L}}$, where $n$ is a loop fugacity (weight),
$N_{L}$ is the number of loops, and $G$ denotes all loop configurations. Like
the FPL model on the square and honeycomb lattices, $n = 2$ for the 4-8
lattice, but unlike these cases the 4-8 lattice is not critical \cite{FPL4-8}.
This is again consistent with our demonstration that long-range order is
present at finite temperatures. Finally, the 3-bond coloring model on the
4-8 lattice is also equivalent to a 3-vertex coloring model on the
square-kagome lattice \cite{triangular}, and thus the $q = 3$ Potts model
on the square-kagome lattice is equivalent at zero temperature to the $q = 4$
Potts model on the Union-Jack lattice.

For a full understanding of the finite-temperature transition and the
partially ordered ground state, we turn to the thermodynamic quantities
extracted from the partition function. For a heuristic understanding of
the partially ordered state, we begin with the entropy. The $q = 3$
Potts model on the diced lattice [Fig.~1(b)] is the prototypical model
in this class for a ground state with partial order (Ref.~\cite{Diced} and
references therein). As above, if the minority sites order, the majority
sites would have two remaining degrees of freedom (d.o.f.s), giving an
entropy per site $S^D_0 (0) = 2 \ln 2 / 3 = 0.462098$. The entropy we
compute is shown in Fig.~4, and its low-temperature limit is $S^D (0)
 = 0.473839$. This minor deviation indicates that an ideal A-sublattice
order is rather close to the true ground state, with only small contributions
from states of imperfect A order. This partial order is destroyed at $T_c$ by
thermal fluctuations, and at high $T$, where all sites may explore all three
d.o.f.s, the entropy approaches $\ln 3$.

\begin{figure}[t]
\includegraphics[width=8.1cm]{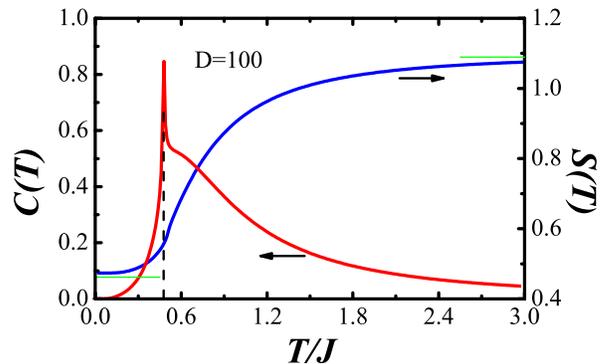}
\caption{(color online) Solid lines: entropy $S(T)$ and specific heat $C(T)$
for the $q = 3$ Potts model on the diced lattice, calculated with $D = 100$.
The dotted line marks the result of Ref.~\cite{Diced}. Thin, green lines mark
the low- and high-$T$ entropy bounds. Dashed lines: results for the $q = 4$
Potts model on the centered diced lattice. }
\vspace{-0.3cm}
\label{ulpmf4}
\end{figure}

Analogous considerations for the Union-Jack lattice give again a maximum
entropy if the high-coordination (minority) sites order. This order may
involve sublattices A, B, or both A and B simultaneously. In the last case,
if $\sigma_A = 0$ and $\sigma_B = 1$, the d.o.f.~in $\sigma_C = 2,3$ yields
a total of $2^{N/2}$ states. If only A sites are ordered ($\sigma_A = 0$,
$\sigma_B, \sigma_C = 1, 2, 3$), the B and C sublattices form a $q = 3$
Potts model on the decorated square lattice. Let the zero-temperature
entropy of this model be $S^{\rm DSL} (0) = \ln \zeta$, then the number
of states with partial order on A only is $\zeta^{3N/4}$, and these will
dominate the total if $\zeta > 2^{2/3} = 1.587401$. We have calculated
separately the value $S^{\rm DSL} (0) = 0.561070$, whence $\zeta = 1.752547$.
Thus the ground state is indeed composed primarily of states with one ordered
sublattice. The discrepancy between $3 S^{\rm DSL} (0) / 4 = 0.420802$ and
our result [Fig.~3(a)], $S^{\rm UJ} (0) = 0.430999$, can be ascribed to
states with neither A nor B order. The high-$T$ entropy approaches $\ln 4$.

We now return to the specific heat and to the phase transition. Results
for the $q = 3$ model on the diced lattice are shown in Fig.~4. The
transition point may be obtained with high precision and we find $T_c
 = 0.505(1)$, a value consistent with the alternative approach used in
Ref.~\cite{Diced}. The peak feature is much more pronounced than
in Fig.~3. This reflects the relative lack of competition between
different types of partially ordered state in the diced lattice (A order
only) as compared to the Union-Jack lattice (A or B order). We note that
both transitions occur near the peak of the specific-heat curve at
$T \sim J/2$, which is the characteristic energy scale of the system,
and we suggest that this behavior is generic: a finite-temperature
transition must occur at a value $T_c/J \sim O(1)$, and cannot occur
arbitrarily close to $T = 0$.

\begin{figure}[t]
\includegraphics[width=8.1cm]{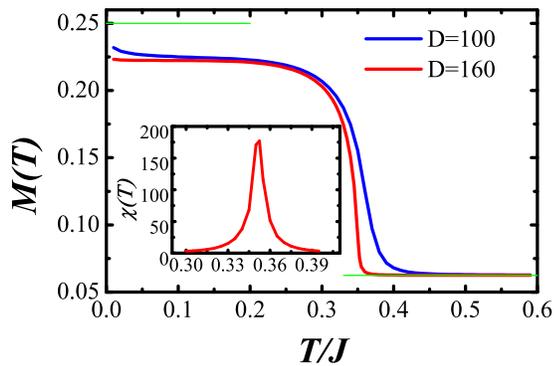}
\caption{(color online) Magnetization $M(T)$ of sublattice A for the $q = 4$
Potts model on the Union-Jack lattice, calculated for an applied field of
0.000025. Thin green lines indicate the low- and high-$T$ bounds. Inset:
magnetic susceptibility $\chi(T)$.}
\label{ulpmf5}
\end{figure}

The other thermodynamic quantities we illustrate here are the magnetization
and susceptibility. By considering finite fields $H$ in the Hamiltonian of
Eq.~(\ref{ehpm}), we deduce the spontaneous magnetization on sublattice A
of the Union-Jack lattice [red circles in Fig.~1(a)], $M = \sum_{i\in A}
\delta_{\sigma_{i=0}}/N$, from the expression
\begin{equation}
 M = -\frac{\partial }{\partial H} \lim_{N \rightarrow \infty} \frac{1}{N}
\ln Z,
\end{equation}
computed with a very small external field ($H = 0.000025$). The results
(Fig.~5) show a clear step around $T_c$. The high-$T$ limit of $M$ for a
completely disordered state is $M^{\rm UJ}(\infty) = \frac{1}{4} \times
\frac{1}{4} = \frac{1}{16}$. The low-$T$ limit for a state of perfect
A-sublattice order would be $M_0^{\rm UJ} (0) = \frac{1}{4}$, and in fact
this value must be obtained for temperatures below the energy from the
applied field. However, it is clear from our entropy calculation that the
ground state is not one of perfect A order. We find that the magnetization
exhibits a related zero-temperature deviation, tending towards the value
$M^{\rm UJ} (0) = 0.2232$. This deviation is somewhat smaller for the diced
lattice, where the ideal magnetization would be $\frac{1}{3}$ and our
numerical result is $0.3192$. Returning to the transition, our calculations
show the same behavior as in Fig.~3(b), that $T_c$ falls slowly with increasing
$D$. The best qualitative indication of the transition is provided by the
magnetic susceptibility, $\chi = \partial M / \partial H$, which has a
robust peak (inset, Fig.~5). The critical exponents of these quantities
may also be computed by the same techniques, but the highly numerically
demanding task of obtaining adequate precision remains in progress.

When the discussion of 2D antiferromagnetic Potts models is framed in terms
of $q_c({\cal L})$, the diced lattice is the only known system with an
average $z$ of 4 but $q_c > 3$. The Union-Jack lattice is now revealed as
the only system known with an average $z$ of 6 but $q_c > 4$. In fact this
is the largest value known for any planar lattice. Phase transitions in
different Potts models belong in general to different universality classes.
While the universality class for the $q = 4$ Potts model on the Union-Jack
lattice remains under investigation, the partial order we find breaks both
the 4-state (Potts) symmetry and the (Ising-like) sublattice symmetry
between A and B.

We end this discussion with the logical extension of our analysis. The 11
regular lattices (all sites equivalent) obtained by tiling the plane with
regular polygons are known as Archimedean. Among their dual lattices, three
are regular and the other eight are irregular. This is the set of Laves
lattices, which includes the diced and Union-Jack lattices. We propose
that for each Laves lattice with an integral average coordination number,
there exists a Potts model with integral $q$ that would feature a
zero-temperature transition on a regular lattice of the same coordination,
but has a finite-temperature transition, to a state of partial order, on
this irregular lattice. As an example we cite the $D(4,6,12)$ lattice, or
centered diced lattice, shown in Fig.~1(c). This tripartite lattice has
an average coordination $z = 6$. Indeed we find (Fig.~4) that a $q = 4$
Potts model on this lattice has a very robust finite-temperature phase
transition to a state of predominantly A-sublattice order. Demonstrating
the existence of the same physics on a third lattice in this class very
strongly reinforces our proposal.

To conclude, we have demonstrated the existence of a previously unknown,
finite-temperature phase transition in the $q$ = 4 Potts model on the
Union-Jack lattice. This establishes the essential property that the
presence of inequivalent sites, leading to a nontrivial entropy, drives the
finite-temperature transition and confers unusually high values on $q_c$.
We find this type of transition on other irregular lattices in 2D. Our
analysis underlines the utility of tensor-based numerical methods in
investigating the physics of classical statistical mechanical models.

We thank Y.-J. Deng for helpful discussions. This work was supported by
the NSF of China under Grant No.~10874244 and by the National Basic
Research Program of the MoST of China.

\end{document}